# A Guided Tour Study of the Untidy But Inspirational PIM of Visual Artists

**Hellmich, Helene** — Humboldt-Universität zu Berlin, Germany | h.hellmich@hu-berlin.de

**Dinneen, Jesse David** — Humboldt-Universität zu Berlin, Germany | jesse.dinneen@hu-berlin.de

**ABSTRACT**

While all individuals deal with increasingly large amounts of digital information in their everyday lives and professionally, prior works suggest visual artists have unique information management practices and challenges. This study therefore examined the personal information management (PIM) practices and challenges of six practising visual artists using guided tours and short interviews. It was found that the visual artists had some unique practices connected to their strong emphasis on serendipity, inspiration, and visual dimensions of information. Like non-artists, the participants faced challenges across all phases of PIM, chiefly an excess of information and fragmented organisation, and they found it especially hard to assess how personal and valuable their information could be. After characterising this rarely discussed PIM demographic, we draw on the findings to provide concrete recommendations for artists doing PIM, for information and cultural heritage institutions, and for designers of PIM software.

## INTRODUCTION

Today individuals deal with increasingly large amounts of information (Vitale et al., 2020) including paper-based and digital *personal information*. This is also true for visual artists, defined here as professional creators of visual art (i.e. not of other art forms like music or dancing). Personal information management (PIM) examines how people keep, organise and re-find their personal information and has been described as "one of life's essential skills" in today's information society (Jones et al., 2017, p. 3584; Whittaker, 2011). The PIM of visual artists has only recently been examined in a small number of preliminary studies (reviewed below), which together suggest their practices and challenges may be unique and thus potentially elucidating for PIM research and tools more generally. It is also valuable to learn about how artists approach (or neglect) PIM and their struggles therein, because their personal information collections, *if preserved,* could become cultural heritage worth saving for future generations (Krtalić et al., 2021; Post, 2017a).

The aim of this research is to discover more about the PIM of visual artists and the challenges they face regarding the management of their personal information. These insights might improve the visual artists' current day-to-day interactions with their personal information and address any PIM challenges they face, as well as guiding collaborations with memory institutions ultimately leading to the preservation of visual artists' personal information collections over the long-term. Additionally, the findings of this study can inform the development of PIM tools to match the requirements and practices of visual artists. The study in particular is concerned with the digital personal information of the participants that is related to their work as artists and that they manage on their computers, and seeks specifically to characterise their PIM practices and challenges (and especially their potential novelty).

As little is known about artists' PIM, we employed an exploratory and qualitative approach, conducting interviews and guided tours with six practising visual artists with the goal to characterise their PIM and generate recommendations for individuals, information institutions, and PIM system designers and researchers.

## LITERATURE REVIEW

Here we briefly review PIM practices and challenges before outlining the gaps in knowledge around artists' PIM.

*PIM and PIM practices* -- PIM is "the practice and the study of the activities a person performs in order to acquire or create, store, organize, maintain, retrieve, use, and distribute information in each of its many forms... as needed to meet life's many goals" (Jones et al., 2017) and today increasingly happens in the digital realm (Vitale et al., 2020). In the foundational PIM literature, both Jones (2007; Jones et al., 2017) and Whittaker (2011) propose a framework (or mode) for PIM activities describing a life-cycle of personal information, but whereas Whittaker (2011) names the three PIM stages *keeping, managing* and *exploiting,* Jones, (2007) calls them *keeping, managing* (or *meta-level activities*) and *(re-)finding*. It has been argued that these models are roughly equivalent because exploitation and (re-)finding are both describing how personal information can be retrieved and used (Dinneen & Julien, 2020). Since *re-finding* is less likely than *exploiting* to have an alternate interpretation in any given use and a narrower subset of management practices has been examined here, this study uses the terms **keeping**, **organising** and **re-finding**.

The keeping stage includes actions towards any newly encountered (Jones, 2007; Jones et al., 2017) or self-created information (Whittaker, 2011) as well as addressing whether the information should be kept or deleted (Vitale et al., 2019; Whittaker, 2011). Organising activities include ways in which individuals structure their files or folders (Dinneen et al., 2019), tidying up information, for example, using filing, piling or mixing strategies (Malone, 1983; Trullemans & Signer, 2014) or recommended PIM practices like storing information by project, using archive folders (Jones et al., 2015) and setting up reminders. Finally, re-finding personal information entails finding information via search or navigation to then use (or *exploit*) it (Jones, 2007; Jones et al., 2017; Whittaker, 2011). In practice, PIM activities may extend across categories (Jones et al., 2017); for example, deleting a file could occur in the keeping stage (Vitale et al., 2018) or in the organising stage (Dinneen & Julien, 2020) of the PIM activities.

Although some PIM practices have been examined frequently (e.g. tagging, a kind of organising), there is no list of common PIM practices in the literature. This is perhaps understandable since PIM studies tend to focus on particular factors like emotions (Whittaker & Massey, 2020), formats (e.g. files; Dinneen & Julien, 2020), tasks (e.g. keeping; Vitale et al., 2018), tools (e.g. emails; Whittaker & Sidner, 1996) or demographic groups (e.g. knowledge workers; Alon et al., 2019). Reasons for this may include that PIM practices are determined by individual differences (Alon & Nachmias, 2020a; Boardman & Sasse, 2004), PIM practices are changing with the introduction of new technologies (Alon et al., 2019) and most PIM studies focus on particular aspects of PIM behaviour or specific practices (Alon & Nachmias, 2020b).

*PIM challenges* -- Not only is there generally more information available than ever before (Johnson, 2014), individuals also deal with large and growing amounts of personal information (Vitale et al., 2018). The accumulation of information makes all stages of PIM (selecting, organising and re-finding) more difficult and might lead individuals to feel stressed (Vitale et al., 2019). Determining the value of information items can be difficult (Whittaker, 2011) and individuals struggle to decide if they should keep or delete information (Boardman & Sasse, 2004; Vitale et al., 2018; Whittaker, 2011). This often leads to a tendency to keep information by default (Boardman & Sasse, 2004; Vitale et al., 2019), which in turn makes retrieving information more difficult and time consuming. If individuals do not have the time to process all the information they encounter or use it meaningfully, this experience is called information overload (Bawden & Robinson, 2020; Whittaker, 2011). Information overload can affect individuals negatively by deteriorating their health, making them less efficient, reducing their ability to make decisions, be productive, innovative and even creative (Bannon, 2006; Bawden & Robinson, 2020; Johnson, 2014).

Another aspect that makes the complex task of PIM difficult, especially in the digital realm, is that information can be experienced and stored on various devices as well as online platforms, thus forming ecosystems of personal information (Vertesi et al., 2016; Vitale et al., 2018). The phenomenon of storing information in various locations or applications, each using different organising schemes, is known as information fragmentation (Jones et al., 2017). Information fragmentation affects all three stages of PIM (keeping, organising, re-finding) and makes it harder for an individual to have an overview of their personal information (Jones et al., 2017; Vitale et al., 2018, 2019).

As individuals accumulate personal information throughout their lifetime, this information has to be managed not only in the short term, but also over long periods of time (Jones et al., 2016), including long-term project management (Copic Pucihar et al., 2016). Individuals keep personal information for the use beyond the immediate future (Bass, 2013) or to create a legacy (Kaye et al., 2006). There seems to be a tension between the increasing amounts of data that individuals keep (Vitale et al., 2019) and what they would like to share with future generations (Gulotta et al., 2013), ideally selecting only remarkable information to keep (Lindley et al., 2013; Vitale et al., 2019). Visual artists fall into one of the demographic groups whose personal collections might be valuable as cultural heritage and not only to the individuals personally (Krtalić et al., 2021; Post, 2017a). Previous research investigated the personal collections and PIM behaviour of notable people, but this research did not focus on digital collections (Krtalić et al., 2021). The personal collections of less well-known individuals, like emerging artists who might become famous or local artists, might equally be considered culturally significant in the future and thus should be kept (Post, 2017a). Knowledge about the PIM of visual artists can inform institutions that might potentially be interested in keeping this information, raise the awareness of artists regarding their own PIM practices and challenges and guide the future development of PIM tools for visual artists (Krtalić et al., 2021; Post, 2017a).

*Visual artists and PIM* -- Although PIM tends to have an individual focus, PIM practices and challenges are often influenced by common demographic features, such as occupation, and such groups are therefore examined in PIM studies despite the usual focus on individuals. For example, PIM studies have examined teachers (Diekema & Olsen, 2014), knowledge workers (Alon et al., 2019) and immigrants (Krtalic, 2021), among others. The demographic group of visual artists, however, has only been the focus of one recent study addressing the personal collections of writers and visual artists in New Zealand, which explored how they perceive the value of their personal collections, their collection management practices and challenges, and the influence these phenomena may have on the possible future use of such collections (Krtalić & Dinneen, 2022). Among the findings were that the participants saw value in their personal collections and used idiosyncratic management practices (Krtalić & Dinneen, 2022).

The information behaviour of artists has been described as distinctly different from other demographic groups (Hemmig, 2008; Mason & Robinson, 2011). They were found to seek especially visual information for inspiration and use extremely idiosyncratic sources of inspiration, including intangible information (Hemmig, 2008, 2009; Mason & Robinson, 2011). Artists have been reported to potentially call anything a source of inspiration, often drawing on information sources that have no direct relationship to the art world (Hemmig, 2009; Mason & Robinson, 2011). Furthermore, they are described as information gatherers who use browsing to find information (Hemmig, 2008, 2009; Post, 2020). Browsing makes unintentional information encountering (serendipity) possible. Serendipity has been studied in information behaviour (Erdelez, 2005; Erdelez & Makri, 2020; Liu et al., 2022; Mason & Robinson, 2011) and, for example, as a design principle (Reviglio, 2019). Further, a connection between finding inspiration and serendipity has been made (Peterson, 2020).

Although not explicitly using a PIM perspective, a recent PhD thesis examined the value and management of digital information of visual artists in the UK, focusing on the wider context of what information practices are involved in art making and the value of digital information in artistic practice (Molloy, 2021). As might be expected, the use of digital information was found to be fundamental to contemporary art practices. However, the participants lacked the relevant digital skills and expertise, which were described as invisible skills and invisible labour (Molloy, 2021).

Previous research also examined the long-term preservation of new media artists and the collaboration between a local artist and an institution to preserve his personal information collection over the long term (Post, 2017b, 2017a). From these studies, we know that media artists keep original digital artworks and digital documentation of their artworks. They use their personal archives for various professional activities, including applying for grants, making portfolios or artist statements and addressing estate concerns. At the same time, new media artists believed that making artwork was more important than thinking about preserving it for future generations (Post, 2017b).

*Summary* -- Disparate studies have begun to indicate that artists' PIM practices and challenges may be useful to learn about and support, but such phenomena have not been systematically investigated. Analysing the PIM of visual artists will help to understand how they keep, organise and use their personal information as well as what challenges they face. The importance of visual information and the heterogeneity of their information sources described above (Hemmig, 2009) might influence visual artists' behaviour or require them to employ particular PIM practices. Previous research on visual artists gives almost no insight into the handling of contemporary, digital, PIM-related items and it has been noted that further research is needed to investigate the personal archives of artists (Post, 2017b). Beyond their use for the individuals themselves, the personal information collections of visual artists could potentially be valuable for others (Post, 2017a). Exploring visual artists' PIM practices and challenges thus has the potential to support artists and institutions in preserving personal information collections, raise artists' awareness regarding PIM, and inform the development of PIM tools.

## METHODOLOGY

### Research objectives and questions

Following the knowledge gaps identified above, the objective of this study is to examine the PIM practices of visual artists related to their work and identify what PIM-related challenges they may face. We thus pose the following research questions (RQs), with a specific concern for the digital personal information collections that relate to artists' work (i.e. we examine no paper-based personal information nor information unrelated to artistic practice).

RQ1: Which PIM practices do visual artists use?
RQ2: What are PIM challenges for visual artists?

To answer the first question, it will be examined what PIM practices artists employ, and which of those are unique or common to non-artist PIM. To evaluate uniqueness, in lieu of an existing list of PIM practices, a list of known practices was compiled from existing literature; this list can be found in the Discussion section as codes (Practices shared by non-artists) in Table 3. Sources for the list included prior PIM publications that (a) give a comprehensive overview of PIM (Jones et al., 2017; Whittaker, 2011) or (b) collect and name a large number of practices regarding digital PIM (Alon et al., 2019; Alon & Nachmias, 2020b; Boardman & Sasse, 2004; Jones et al., 2015). Each practice was assigned to one of the three common PIM stages discussed above (keeping, organising, and re-finding).

### Data collection

We employed a qualitative approach to identify possible practices and challenges and elicit further information about the relatively unknown phenomenon of artists' PIM (Connaway & Radford, 2016).

*Sample and recruitment* -- A purposive sample of six participants was used in this study. Although non-probabilistic sampling often relies on data saturation to determine the necessary number of participants (Guest et al., 2006), it is unlikely that data saturation would be attained in this study because of the nature of the participant group and of PIM itself; rather, the individualistic nature of PIM (Boardman & Sasse, 2004; Jones et al., 2015) and especially the uniqueness of artists with idiosyncratic information needs and types (Hemmig, 2009; Mason & Robinson, 2011) are among the study's premises and are suggested by the research reviewed above. A sample size of six participants has been described as a conventional baseline for such studies (Helfferich, 2009, p.175) and recent meta-research indicates that in exploratory, qualitative studies "typically 6-7 interviews will capture the majority of themes" (Guest et al., 2020, p.13).

Participants were recruited through the authors' connections to the local visual artist community (i.e., recruitment via gatekeeper). An important consequence was that some level of trust had already been established between the authors and the participants before the start of the guided tour, which is recommended (Thomson, 2018) since during guided tours participants' private spaces and personal information are exposed. Participants were selected only if their primary profession is that of a practising visual artist (e.g. exhibiting and/or receiving grants) and they manage information themselves (e.g. without the help of an assistant). Although it was not a study goal to generalise the findings across all artists, variation was sought among participants practising different art forms (e.g. sculpture, drawing, painting, video) to avoid the findings being solely about practitioners of one art form and to instead enable examining a breadth of practices and challenges within the sample.

*Participants* -- A summary of the participants' demographic and criteria-related data can be found in Table 1. All participants were working in Germany. The participants included three men, two women, and one non-binary person, each practising different art forms. All participants showed their laptops during the guided tour, with the exception of P1, who did not own a laptop and showed his desktop computer; all participants used MacOS.

| # | Art form | Gender | Device |
|---|---|---|---|
| P1 | installation, drawing | man | desktop |
| P2 | installation with video | man | laptop |
| P3 | sculpture | non-binary | laptop |
| P4 | drawing | man | laptop |
| P5 | conceptual, documentary video and sound | woman | laptop |
| P6 | painting | woman | laptop |

**Table 1. Summary of study participants.**

*Procedure* -- The guided tour is a technique that has been applied to research in physical (Malone, 1983; Petrelli & Whittaker, 2010) and digital personal information spaces (Barreau, 1995; Boardman & Sasse, 2004; Vitale et al., 2018), wherein the participant leads the researcher through the physical or digital space in question while sharing thoughts and stories in a manner similar to the think-aloud protocol, thus leveraging a visual method (Hemmig, 2009) to elicit a description of the collection and related processes and perceptions (i.e. with the benefit of examples and not relying solely on memory) and allowing participants to show only what they wish (Thomson, 2018). The guided tours were conducted in October 2021 following a protocol and the guidelines of Thomson (2018; e.g. participants were assured their PIM would not be judged) and included only the digital spaces of participants' desktop or laptop computers where personal materials related to their artistic work were stored. Two pre-tests suggested the approach was effective for collecting the necessary data and would take approximately 75 minutes per participant, which falls between the times reported in comparable guided tours (e.g. 45 minutes; Vitale et al., 2018, and 90 minutes; Bergman, 2013). Five of the guided tours were conducted in person and one remotely (i.e. with screensharing). As a guided tour generally reflects the current state of a participant's personal information and, to some extent, their past practices, post-tour interviews were used to ask about the future of the collection. Sessions of one tour and interview each were video recorded and lasted between 47 minutes and 75 minutes.

## Data analysis

The session recordings were transcribed and thematically analysed in MAXQDA Plus 2020 (20.4.1) until the themes below were identified (Braun & Clarke, 2006). Deductive codes (the practices and challenges collected during the literature review) and inductive codes (practices or challenges mentioned by participants but not found during the review, hereafter *unique practices and challenges*) were identified successively. The resulting themes are summarised with example codes in Table 2.

|  | Themes | Example codes | Description |
|---|---|---|---|
| PIM practices | Keeping everything | selecting, backing-up | What information was used and if they kept or deleted information. |
|  | Not tidying up | piling, tagging | How and where information was kept by visual artists and if they tidied up their information. |
|  | Using everything as inspiration | navigating, searching | How the visual artists retrieved information. |
| PIM challenges | Too much information | information overload, experiencing time constraints | Challenges experienced due to the amount of information kept. |
|  | Allowing too much chaos | information fragmentation | Challenges experienced due to information fragmentation. |
|  | (Future) value of information | determining future value, "finding it later" | Challenges experienced by the visual artists regarding the value of their information. |

**Table 2.** Themes resulting from thematic analysis of interview and guided tour data (Braun & Clarke, 2006).

## RESULTS

### PIM practices of visual artists

The PIM practices of the visual artists are described within three themes: *Keeping everything, Not tidying up* and *Using everything as inspiration*. The practices of individual participants were all very different from one another and consensus about any particular practice was rare. These cannot be fully reviewed here and instead will be summarised in favour of showing the overall trends.

*Keeping everything* -- The visual artists from this study created, received or collected digital information. Most of the documents, kept by the visual artists were self-created original artworks (P2, P5) like videos, documentary video- and audio-recordings or supportive material (all participants), especially digital reproductions of physical artworks. The digital reproductions were often compiled to serve as a basis to create various other types of documents (applications, portfolios, websites, newsletters, catalogues) or new artworks (books, films). The artists used and continuously accessed new and old digital materials to create these compilations. The participants mainly worked with images: *I operate with images pretty much all the time* (P1) and used images in many different formats, as each use case required different resolutions, file formats (e.g. RAW, tiff, jpg, psd) or different editing of the images. Especially P1 and P2 collected information extensively, mostly in the form of inspirational material from the web. Overall, 5 of the 6 participants preferred to keep digital information and avoided deletion even if the information was not useful. One participant even created an extra folder called "discarded" for materials that she had rejected in her selection process: *if something is in there that is good then I select it and put it into another folder and the rest stays there... I don't delete it. ... Those are all the repros that have been discarded.* (P6)

*Not tidying up* -- Neither computer desktops nor folders of the participants were organised clearly or consistently. All but one participant (P3) used a rather chaotic organising system: *It is just a huge chaos.* (P5). Five of the six did not use coherent naming schemes, nor clear folder structures. They either mixed different information together in one project folder, saved information items in the wrong folder or kept duplicates of the same information in many different places. Regarding folder structure, the most important practice of the participants was to sort information by project in an "art works" folder and an additional "grant applications" folder. Four participants keep all their projects (sometimes more than 50) sorted by name in a single folder. The participants did not use archive folders for older projects. Only P6 had a folder called "old applications" that could be described as an archive folder.

All of the participants mentioned that they felt that their desktop was untidy or referred to having the impression of it being chaotic or messy: *The desktop is … kind of unsorted and there's pretty much only stuff which I forgot about.* (P4) Figure 1 shows that P1 stored almost his entire information directly on the desktop, which he navigated like a map. The strategy of placing information in specific places on the desktop was also used by other participants.

*Sometimes there are these moments in which I start to arrange things towards the top or more into the middle and everything else to the bottom or the side, so that I can access certain folders very quickly for a meeting ...* (P2)

In contrast, as visible in Figure 1, P3 kept only current information on his desktop: *I have the ... desktop pretty empty actually... It just depends what I'm working on at the moment.* (P3), a practice that was also used by other participants: *on the desktop I really have everything current, so, more or less current things that I kind of want to access quickly.*

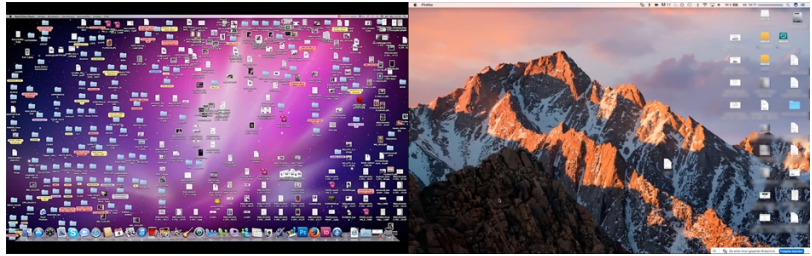

Figure 1: Screenshots of the desktops of P1 (left), P3 (right). File names have been blurred.

Five of the six participants did not tidy up their digital personal information regularly. Two of the participants explained that they did not have the time or motivation to clean up: *most of the time it's like, that I, when I have finished something, then I'm just glad to be done and … I don't tidy up afterwards. It's like after cooking, if I don't do the dishes.* (P4) P5 noticed a difference between tidying up her artworks and "grant applications": *like as if it would give me a better framework, if someone else directs me from the outside. So apparently, I am handling these applications more orderly than my artistic work.* (P5)

To be able to access information more quickly, participants created reminders, most often in the form of screenshots: *if I'm on the internet and I see things, I quickly take a screenshot, then they all accumulate in the top right corner.* (P1) or by leaving windows open (used by half of the participants). Screenshots were also used by all artists for inspirational purposes: *these are like stimulations or after research on certain materials, which I then save there.* (P2) The screenshots were usually kept for their visual value and not as a reference to the original information, like bookmarks. Colour tags were used by all artists to mark selected images or to highlight projects that were recent, important, or in progress: *the really important things, they then always have a colour like this. Or the things … I access the most, have some colour and always a different one, so that, so that I can access them more quickly.* (P6)

*Using everything as inspiration* -- The method preferred by participants for re-finding personal information on their computers was navigation and some even *never* used search (P4, P6), even if navigation was not successful immediately: *if I had somehow been looking for these photos, I would have, I think, first gone here into the works and then here in [name of folder] … they're not in there, so I guess I have to look somewhere else.* (P6) Four of the participants used the icons of the image preview to locate the correct file, stressing the role of their visual memory in the search process: *if I see those in small, then I know right away which file it is. So, I know, okay, this is from that exhibition.* (P1) Other participants used a mix of navigation and search. Some participants explicitly stated that instead of deliberately searching for information, they preferred to be inspired and enjoyed encountering information by chance as part of their creative process. P1 even saw his whole organising system as a way to accidentally discover information and regarded all his digital materials as potential sources of inspiration: *And there is only this one connection with a very specific moment, with a very specific situation, an image and the connection to that text. And that is something that one has to rediscover first of all.* (P1)

### PIM challenges of visual artists

The PIM challenges of the visual artists are described within three themes: *Keeping too much information, Allowing too much chaos* and *(Future) value of information*.

*Keeping too much information* -- Four of the six participants stored large amounts of digital personal information. They were also aware that the accumulation of information was problematic, yet they found it difficult to delete anything: *it's always like that, things easily come to you, but then then it's difficult to let them go.* (P4) Accumulating too much information was overwhelming for the participants: *this potentially available material, I mean, it's also totally overwhelming.* (P1) One participant imagined that even without adding any new material, he would already have enough information to work with it for "years and decades" (P1). On the other hand, keeping a lot of clutter prevented the artists from concentrating on their tasks: *and that's why I would like to leave behind intermediary steps or bad quality and so on, to be able to concentrate better, so that I am not as distracted from what is essential.* (P4)

*Allowing too much chaos* -- Information fragmentation occurred across devices, platforms, project spaces, formats and versions and affected participants to differing degrees as well as provoking negative emotional responses. Different versions of information items were kept abundantly by participants and versions were often hard to distinguish from duplicates: *First of all, there are two different sizes, then I updated them because new ones were added … there are ... different versions of the identical looking portfolio.* (P1) The final versions of documents were usually not marked and therefore could not be found easily. One participant (P2) reported that keeping her information in disorder was making her anxious and scared: *There is just such a huge chaos and I'm more scared of that*. (P5) Not finding information items could be very frustrating for the artists and caused many negative emotions or confusion: *Some things I can't find again immediately, then I have to look for a while… that's more annoying.* (P1) All of the artists expressed (mostly implicitly) that they thought that there was a right way to do PIM. Usually, they implied that tidiness would be more acceptable. As a result, participants felt bad and judged themselves for not being on top of things, not tidying up or doing PIM 'wrong': *well, look, actually this is also totally idiotic... which is completely stupid.* (P5) Half of the participants felt embarrassed towards the interviewer for being messy, as P1 expressed: *it is already quite an effort to bring myself to show it like this.* (P1)

*The (future) value of information* -- Five of the six participants did not know if any of the information they kept was even valuable at all or which information could be valuable: *I think one would simply have to look at it all, as in: What is there? ... What is even relevant? ... What will even become relevant again?* (P5) Some of the visual artists (P1, P2) even found it presumptuous to claim that their own digital work would be important enough to be kept as a legacy: *it seems that for me it has some sort of ... relevance. But that doesn't automatically mean that it is somehow relevant to somebody else.* (P1) Putting personal collections into order was seen as a prerequisite to making them accessible to others: *to have some sort of order in there, so that it is accessible for other people.* (P5) Identity construction and sentimental reasons were not mentioned by the artists as being important aspects of their digital personal information. Most prevalent among the participants was a practical approach of keeping information to "find it later" (Kaye et al., 2006). Thus, (long-term) preservation was not a pressing concern for the participants: *it's only really for now. So, I don't think about what it will look like in the long term*. (P6)

## DISCUSSION

### RQ1: Which PIM practices do visual artists use?

The visual artists in this study used many common PIM practices and also employed several unique PIM practices that have not previously been described in the literature. As could be expected, many of these unique practices are visual. Another finding that stands out is that some of the participants enjoyed discovering their personal information serendipitously. Table 3 summarises which common PIM practices were applied by the visual artists and Table 4 shows their unique PIM practices. Each is discussed below.

| Frequency | Keeping | Organising | Re-finding |
|---|---|---|---|
| widely used | creating, acquiring (collecting, receiving), keeping / saving, assigning value, backing-up (external hard-drive) | using desktop, using folders and folder structure, naming files & folders meaningfully, sorting by project, mixing, being inconsistent, re-naming, abandoning, creating empty folders, referring to ideal behaviour, accessing quickly, leaving windows open | navigating, searching, remembering |
| used little or just by one | selecting, deleting, backing-up, backing-up (cloud storage) | re-organising, being efficient, filing, piling, tagging, creating to-do lists, using apps & tools, creating to-do lists, bookmarking | |
| not used | | using archive folder | |

**Table 3. Common practices used by the visual artists.**

*Common practices* -- The keeping activities of the visual artists could generally be described as "keep everything", as they rarely deleted information even if it was not useful, which is common for individuals (Sweeten et al., 2018; Vitale et al., 2018; Whittaker, 2011). Most of the personal information examined were self-created images. The organisation of personal information on their computer desktops and in folders was mostly chaotic; participants, for example, did not tidy up their desktops or placed all projects in a single folder in a flat hierarchy, which is atypical but nonetheless done by some non-artists(Dinneen et al., 2019; Dinneen & Julien, 2019). Organising systems were used inconsistently and the personal information was not tidied up frequently, adding further clutter and making it hard to re-find information. The preference for re-finding files by navigating to them (e.g. rather than using desktop search) is consistent with previous studies (Bergman, 2013; Dinneen & Julien, 2020; Malone, 1983). (Jones et al., 2017; Whittaker, 2011)

*Unique practices* -- In total eight unique practices not yet observed in the reviewed PIM literature, were identified as being performed by the visual artists, as shown in Table 4. The unique practices were observed in all three PIM stages (keeping, organising, re-finding) and were all widely used by the participants.

| Keeping | Organising | Re-finding |
| --- | --- | --- |
| Compiling: combining old information (mostly images) to create new projects (e.g. portfolios, publications, Websites, applications) Reproducing: making digital photographic reproductions or scans of art works | Using the desktop like a map (i.e. placing icons in meaningful locations around the desktop) Using tags' colours for visual reminders Making screenshots as visual reminders | Explicitly using visual memory to remember and re-find information Intentionally using information from the personal information collection as inspiration Intentionally creating the conditions for serendipitous information encounters (in addition or in contrast to planned re-finding) |

**Table 4. Unique practices used by the visual artists.**

Perhaps unsurprisingly, many of the practices unique to visual artists were *visual* in nature or concerned *images* specifically. For example, rather than keeping old versions or reproductions purely for documentation purposes, as individuals concerned with long-term preservation of personal collections might do (Krtalić et al., 2021), the artists re-use the items in or as inspiration for new projects. This is arguably a visual version of behaviour observed in teachers who re-use content from their personal collections when creating new material (e.g. lesson plans; Diekema & Olsen, 2011). Some of the artists used their computer desktop like a map, organising files and folders visually in certain areas of the desktop. We know that individuals keep large amounts of images in personal file collections (Dinneen & Julien, 2019), but not necessarily how they are organised and used. Using the desktop like a map also served the participants to re-find information using their visual memory.

The use of web-content by artists has been described previously (Mason & Robinson, 2011; Robinson, 2014), but not the specific use of screenshots as reminders or for inspiration. The visual artists also used colour tags as reminders. Instead of ascribing a meaning to a particular colour, they just used the colour to draw attention to a particular file or folder. For sure, many individuals besides visual artists use colour tags, but this practice has not explicitly been studied in the PIM research. The visual reminders like screenshots and colour tags also functioned as visual shortcuts to re-find information. Visual memory was also used by looking at preview images to identify a particular version of a film or image, a behaviour Bergman (2013) also observed in a graphic designer.

The visual artists regarded their personal information as potential material for inspiration, confirming Hemmig (2009). Some further exploited this by using serendipitous discovery (Erdelez, 2005) within their digital personal information collections instead of searching for particular information. The importance of serendipity has been recognised in the design of information architectures (Reviglio, 2019), but to our knowledge, the accidental encounter of information has not been connected to PIM practices or tools. While serendipitous discovery may be distracting or inefficient during known-item retrievals, it could also be more rewarding or inspiring (Mason & Robinson, 2011; Reviglio, 2019). Perhaps there is a relationship between the preference for serendipity, not deleting personal information, and the unstructured way of keeping artistic projects, thus giving the artists the opportunity to be surprised and inspired by their personal information in their creative process.

### RQ2: What are PIM challenges for visual artists?

The challenges the visual artists faced were all common challenges and no unique challenges were identified. Most challenges were experienced by all of the participants and were related to information overload, information fragmentation, and determining the (future) value of the personal information.

*Overload and fragmentation* -- The increasing complexity of PIM and the resulting challenges have been remarked upon in the PIM literature (Alon et al., 2019). In contrast to knowledge workers, however, the visual artists who participated in this study did not employ any high-level strategies to deal with these challenges. Instead, most of them rather lost their motivation to engage in any PIM activities. Both information overload (Bawden & Robinson, 2020; Jones et al., 2017; Whittaker, 2011) and information fragmentation (Jones et al., 2017; Vertesi et al., 2016; Vitale et al., 2018, 2019) are common PIM challenges and were experienced to varying degrees by all artists. The participants collected large amounts of digital personal information and did not tidy up their information spaces regularly or consistently, as they perceived PIM activities to be a waste of time. Thus, they kept information (mostly images) fragmented across versions, formats and duplicates (Jones et al., 2017; Vitale et al., 2018, 2019) which also made it difficult and time-consuming to re-find information (Whittaker, 2011). Due to information overload, the participants found it hard to make decisions or concentrate and it caused them to feel overwhelmed, anxious, frustrated or lead to information avoidance, which is common (Bawden & Robinson, 2020). Additionally, the participants all expressed feeling bad about their PIM practices and judged themselves for not "doing it right" (Vertesi et al., 2016). Previous literature describes the tension the visual artists felt between actual and ideal PIM behaviour as a common challenge (Alon & Nachmias, 2020b).

*The (future) value of personal digital information* -- The greatest struggle the visual artists from this study faced was determining the value of their digital personal information, which has been described as a common challenge (Whittaker, 2011). The main reason participants gave for keeping information was for future use, but the way they organised it made "finding it later" (Kaye et al., 2006) difficult. Especially the idea of "building a legacy" or estate (Kaye et al., 2006) was a challenging topic for the visual artists, as they did not believe that what and how they were currently storing their digital information would be accessible or valuable to other people. This finding is surprising because previous studies found that visual artists and writers see the long- and short-term value of their personal collections for themselves and for society (Krtalić & Dinneen, 2022) and collections in the family context are valued as records, evidence of an individual's identity and personality and their broader societal value (Krtalić et al., 2021). It would thus seem appropriate that artists should value the artworks they create and the processes around it.

Equally surprising is the fact that the visual artists hardly reported any personal or emotional involvement with their digital information. The literature, however, suggests that digital possessions are regarded as part of one's identity (Chen et al., 2021; Cushing, 2011; Kaye et al., 2006). Interestingly, leaving collections uncurated has been described as one reason why individuals would not view them as part of their identity and self-representation towards themselves (Odom et al., 2014) or others (Krtalić et al., 2021). Additionally, digital information, compared to physical information, may seem less unique and thus less valuable to an individual (Odom et al., 2014). This has been attributed to the "spacelessness" (collections are physically not visible) of digital information (Odom et al., 2014). As the collections become less remarkable (Lindley et al., 2013) it is undesirable to share them with future generations (Gulotta et al., 2013). This behaviour and rationale were observed in the visual artists, endangering their personal information to be lost, even though it might contain valuable insights about the way artists work today, their creative processes or actual digital artworks that should be preserved as part of our cultural heritage.

This would confirm, that the way visual artists approach PIM affects how their personal information is valued (Post, 2017a). If the visual artists were to recognise the uniqueness and broader societal significance in terms of describing "time, place, and events" (Krtalić et al., 2021, p. 170), it could potentially impact the value that they assign to their personal information. This highlights the importance of the artists themselves understanding how essential their PIM practices are in creating valuable information collections for themselves and for others. Molloy, (2021) suggests that this would require a greater appreciation of digital artistic practices in society at large, making this labour visible, as well as close collaboration between visual artists, memory institutions and LIS professionals in these matters.

**Recommendations**

Though the results of this study must be interpreted in their narrow context (see *Limitations,* below), they nonetheless suggest artists may need help selecting relevant information and deleting information to create personal digital collections that are more valuable for themselves and for others. They also seem to need support organising their personal information in a more structured way, and visually, to create a greater accessibility. The possibility of serendipitous discovery, which they favour, should also be taken into account. Addressing these areas would make the PIM practices of visual artists more efficient and consequently reduce their PIM challenges, including determining the value of their personal information and thus reduce the potential loss of important cultural heritage materials. The recommendations aim to address these concerns.

*Recommendations for artists* -- Make tough saving decisions (Alon et al., 2019) and delete more information to make personal information collections smaller and therefore more personal, more accessible and more valuable for the artists themselves and for others and to avoid information overload. To help during the selection process, each project could be sorted into two folders with core and supportive information (Copic Pucihar et al., 2016).

*Recommendations for LIS professionals and cultural heritage institutions* -- Build educational programs for visual artists that cover PIM topics related to digital information, making research more visible and accessible (Molloy, 2021; Post, 2017b). Offer these courses at art schools, institutions that publicly call for submissions or offer grants to visual artists (as these require applications in digital formats) and professional training programs (e.g. in Berlin the Goldrausch project for women artists, https://goldrausch.org/en/, currently providing some technical guidance for making artist websites but no input regarding PIM). This will firstly help provide a platform where artists can exchange their PIM experiences, expelling insecurities towards their own PIM style, or find new PIM practices suitable to their working environment. Secondly, it will share recommendations from PIM research about organising collections (Jones et al., 2015) and the importance of deleting (Hellmich & Dinneen, 2022).

*Recommendations for PIM tool developers* -- Visual artists use screenshots as reminders and icons to re-find information. Therefore, support the flexible use of images in PIM tools, for example by allowing image view of images from several folders simultaneously, making screenshots searchable (visually, e.g. by colour, motif, or reverse image search), and including them in the image manager by default. Support serendipitous discovery of information by (e.g.) selecting and displaying a certain number of random images from the personal information collection or finding a complementary accidental image to a selected one. This could include drawing data from information resources commonly used in creative tasks (Li et al., 2022).

Besides the practical advantages of these proposals, it is advisable not to force tidiness upon visual artists nor reinforce the social stigma that sometimes exists around chaos and clutter. As Copic Pucihar et al. (2016) pointed out, messiness will not just disappear; it is part of PIM and especially visual artists seem to embrace and somewhat utilise it. Some enjoy being surprised by unexpected encounters and inspiration cannot be planned in advance. It thus seems promising to utilise this phenomenon to create PIM tools that support the creative process and help the artists with the PIM challenges they currently face. This will make it easier for visual artists to keep, organise and re-find digital information in the future and preserve their valuable digital material as cultural heritage.

**Limitations**

The main limitation of this study was the small and geographically narrow sample: despite their artistic variety, all six participants were from one region of one country. Thus, soundly interpreting and using the results requires considering that limited sample and that the findings may not generalise. While generalisability was not the study's goal, it is possible that further unique practices and challenges exist, and that other kinds of artists (e.g. performance) have vastly different PIM practices and challenges from the usual demographics studied. It would therefore be desirable to extend the current study to additional participants or recruit participants with different artistic practices, and from non-German and non-European backgrounds. For example, a more diverse sample could be recruited through professional associations or artists organisations (Molloy, 2021). There was also some inherent subjectivity in the identification and naming of the themes (Table 2), which were derived by a single coder and thus may not be fully replicable (however, that the themes align well with the typical PIM categories of keeping, organising, and re-finding suggests they are broadly reasonable). In future studies, using multiple coders would be preferable. Further, all participants used MacOS; while in some regions it is common for artists to use Apple computers, the software used may have an effect on their practices (Dinneen & Frissen, 2020). Finally, this study only investigated local digital collections shared by the artists and predominantly computer files (e.g. rather than email), so the role of additional digital and physical formats in visual artists' PIM remains unclear.

**CONCLUSION**

This study investigated the PIM of visual artists regarding digital personal information that is related to their work and generated a list of digital PIM practices that may be used in future PIM studies. Using the method of the guided tour, data was collected by observing the artists navigate their personal information on their computers. The aim of the study was to gain more knowledge about the PIM of visual artists in the digital realm and the challenges they faced with this. The first research question asked what PIM practices the visual artists use and the second question what challenges they face. It was found that visual artists use common and unique PIM practices, but that they only face common PIM challenges. The unique PIM practices of visual artists included many visual practices. Keeping large amounts of information mostly unorganised could lead to many challenges that were experienced in all three stages of PIM activities (keeping, organising and re-finding), but was also used by some visual artists to make serendipitous discoveries. A major challenge for the visual artists was to determine what information would be valuable in the future. The results informed concrete recommendations for artists, memory institutions and PIM software developers.

This research contributed to the field of PIM by examining the previously understudied demographic group of visual artists. Studying the PIM of visual artists can, first of all, help to set up guidelines or workshops for this particular group, guiding their everyday interactions with their personal information related to their work as visual artists and secondly, guide artists and institutions alike to collaborate on finding appropriate ways of ensuring the (long-term) preservation of valuable personal information of the visual artists to help create their estates and legacies. Finally, the findings of this study can also contribute to the creation of new PIM tools that facilitate the keeping, organising and re-finding of digital personal information for visual artists so that they would be less burdened by its complexity and challenges. All these individual actions together can help to keep the personal collections as cultural heritage and be able to use them in the future.

Future research could examine the PIM of visual artists more holistically, including their paper-based personal information, as well as how visual artists share their information, how they use social media and how they collaborate with others. Additionally, the PIM of visual artists should be studied over the long term to gain more insights into long-term practices, challenges and the visual artists' attitudes towards their digital legacies, including the relationship to their identity. Such insights should enable improved or novel systems and services (Kljun et al., 2015) benefiting visual artists, cultural heritage institutions, and their future work together.